\documentclass[11pt]{article}
\usepackage{graphicx}
\usepackage{subfloat}
\usepackage{epsfig}
\usepackage{amsfonts}   
\usepackage{amssymb}

\begin{document}
\date{}
\title{{\bf{\Large Critical behavior of Born Infeld AdS black holes in higher dimensions}}}
\author{
{\bf {{\normalsize Rabin Banerjee}}$
$\thanks{e-mail: rabin@bose.res.in}}
,$~${\bf {{\normalsize Dibakar Roychowdhury}}$
$\thanks{e-mail: dibakar@bose.res.in}}\\
 {\normalsize S.~N.~Bose National Centre for Basic Sciences,}
\\{\normalsize JD Block, Sector III, Salt Lake, Kolkata-700098, India}
\\[0.3cm]}



\maketitle


\begin{abstract}
Based on a canonical framework, we investigate the critical behavior of Born-Infeld AdS black holes in higher dimensions. As a special case, considering the appropriate limit, we also analyze the critical phenomena for Reissner Nordstrom AdS black holes. The critical points are marked by the divergences in the heat capacity at constant charge. The static critical exponents associated with various thermodynamic entities are computed and shown to satisfy the thermodynamic scaling laws. These scaling laws have also been found to be compatible with the static scaling hypothesis. Furthermore, we show that the values of these exponents are universal and do not depend on the spatial dimensionality of the AdS space. We also provide a suggestive way to calculate the critical exponents associated with the spatial correlation which satisfy the scaling laws of second kind.     
\end{abstract}

\section{Introduction}
Gravitational physics in higher dimensions has been a topic of interest since the advances made in string theory, where most of the efforts have been made at studying theories in space time dimensions greater than $ (3+1) $.  At the same time, black holes in higher dimensional theories of gravity have been found to posses much richer structures than those in four dimensions \cite{myers}. Now a days, it is generally believed that, understanding of the physics of black holes in higher dimensions is essential in order to understand a full theory of quantum gravity. Studying thermodynamic aspects of these black holes is one of the significant issues in this context. 

The study of thermodynamic properties of black holes has been an intense topic of research since the discovery of the four laws of black hole mechanics by Bardeen, Carter and Hawking in the early seventies \cite{bch}. Since then the question regarding various thermodynamic aspects of black holes, specially the issue of phase transition and critical behavior in black holes remain highly debatable and worthy of further investigations. The study of phase transition as well as the critical phenomena in black holes requires an extensive analogy between the laws of black hole mechanics and that of the ordinary laws of thermodynamics. An attempt along this direction was first commenced by Davies \cite{dav} and Hut\cite{hut}, where they studied the phase transition phenomena in $ (3+1) $ dimensional Kerr Newman and Reissner Nordstrom black holes. The investigation of the thermodynamic behavior of black holes in AdS space was first made by Hawking and Page \cite{hp}. Their analysis was strictly confined to the $ (3+1) $ dimensional Schwarzschild AdS space time. However, in recent years, thermodynamics of black holes in AdS space has attained renewed attention in the context of AdS/CFT duality where one can identify a similar thermodynamic structure in the dual conformal theory residing at the boundary \cite{witten}. A novel approach to study and classify the phase transition phenomena in black holes, based on the Ehrenfest's scheme \cite{zeman} of standard thermodynamics, has been recently initiated \cite{bss} and applied extensively to $ (3+1) $ dimensional black holes in AdS space\cite{modak}-\cite{dibakar3}. In this approach one can in fact show that black holes in $ (3+1) $ dimensional AdS space can undergo a second order phase transition in order to attain thermodynamic stability \cite{modak}-\cite{dibakar3}. Like in  $ (3+1) $ dimensions, black holes in higher dimensional AdS space have also been found to posses a stable thermodynamic structure \cite{dibakar2}.  

In ordinary thermodynamics, a phase transition occurs whenever there is a singularity in free energy or one of its derivatives. The corresponding point of discontinuity is known as the critical point of phase transition. In Ehrenfest's classification of phase transition, the order of the phase transition is characterized by the order of the derivative of free energy that suffers discontinuity at the critical points. For example, if the first derivative of the free energy is discontinuous then the corresponding phase transition is first order in nature. On the other hand, if the first derivatives are continuous and second derivatives are discontinuous or infinite then the transition may be referred as higher order or continuous.  These type of transitions correspond to a divergent heat capacity, an infinite correlation length, and a power law decay of correlations near the critical point. 

The primary aim of the theory of phase transition is to study the singular behavior of various thermodynamic entities (for example heat capacity) near the critical point. It order to do that one often expresses these singularities in terms of power laws characterized by a set of \textit{static} critical exponents \cite{stanley1}-\cite{stanley2}. Generally these exponents depend on a few parameters of the system, like, (1)the spatial dimensionality ($ d $) of the space in which the system is embedded, (2)the range of interactions in the system etc. To be more specific, it is observed that for systems possessing short range interactions, these exponents depend on the spatial dimensionality ($ d $). On the other hand for systems with long range interactions these exponents become independent of $ d $, which is the basic characteristic of a \textit{mean field theory} and is observed for the case $ d>4 $ in usual systems. It is interesting to note that the (static) critical exponents are also found to satisfy certain \textit{thermodynamic scaling laws}\cite{stanley1}-\cite{stanley2} which apply to a wide variety of thermodynamical systems, from elementary particles to turbulent fluid flow. Such studies have also been performed, albeit partially, in the context of black holes\cite{cr1}-\cite{cr17}. In this paper we provide a detailed analysis of these issues and show how the study of critical phenomena in black holes is integrated with corresponding studies in other areas of physics adopting a mean field approximation.

For the past two decades gravity theories with Born-Infeld action have garnered considerable attention due to its several remarkable features \cite{prdw}-\cite{jhepcai}. For example, Born-Infeld type effective actions, which arise naturally in open super strings and D branes are free from physical singularities. Also, using the Born Infeld action one can in fact study various thermodynamic features of Reissner Nordstrom AdS black holes in a suitable limit.  During the last ten years many attempts have been made in order to understand the  thermodynamics of Born-Infeld black holes in AdS space \cite{fernandogrg}-\cite{olivera}. In spite of these efforts, several significant issues remain unanswered. Studying the critical phenomena is one of them. In order to have a deeper insight regarding the underlying phase structure for these black holes  one needs to compute the critical exponents associated with the phase transition and check the validity of the  scaling laws near the critical point. Although an attempt to answer these questions has been commenced very recently \cite{dibakar4}, still a systematic analysis of critical phenomena for higher dimensional Born-Infeld AdS (BI AdS) black holes is lacking in the literature.  A similar remark also holds for the higher dimensional Reissner Nordstrom AdS (RN AdS) black holes \cite{rn1}-\cite{rn3}.   

In this paper, based on a canonical framework, we aim to study the critical behavior of charged black holes taking the particular example of Born-Infeld AdS (BI AdS) black holes in $ (n+1) $ dimensions. Results obtained in the above case smoothly translate to that of the $ (n+1) $ dimensional RN AdS case in the appropriate limit. The critical points of the phase transition are characterized by the discontinuities in heat capacity at constant charge $ (C_Q) $. We compute the static critical exponents associated with this phase transition and verify the scaling laws. These laws are found to be compatible with the \textit{static scaling hypothesis} for black holes \cite{stanley1}-\cite{stanley2}. Finally, taking the spatial dimensionality of the space time as $ d=n $, we also provide a suggestive argument in order to compute the critical exponents associated with the spatial correlation.   

Before we proceed further, let us briefly mention about the organization of our paper. In section 2 we make a qualitative as well as quantitative analysis of the various thermodynamic entities for $ (n+1) $ dimensional BI AdS black holes in order to have a meaningful discussion on their critical behavior in the subsequent sections. In section 3 we study various aspects of the critical phenomena both for the BI AdS and RN AdS black holes in $ (n+1) $ dimensions. Finally, we draw our conclusion in section 4.

\section{Thermodynamics of charged black holes in higher dimensional AdS space}
In this section we compute the essential thermodynamic entities both for the BI AdS and RN AdS black holes that is required in order to explore the critical behavior of these black holes.

The action for the Einstein- Born-Infeld gravity in $ (n+1) $ dimensions $ (n\geq 3) $is given by, \cite{cai},
\begin{equation}
S= \int d^{n+1}x\sqrt{-g}\left[ \frac{R-2\Lambda}{16\pi G}+L(F)\right] 
\end{equation}
where,
\begin{equation}
L(F)=\frac{b^{2}}{4\pi G}\left( 1-\sqrt{1+\frac{2F}{b^{2}}}\right) 
\end{equation}
with $ F=\frac{1}{4} F_{\mu\nu}F^{\mu\nu} $. Here  $ b $ is the Born-Infeld parameter with the dimension of mass and $ \Lambda(=-n(n-1)/2l^2) $ is the cosmological constant. It is also to be noted that for the rest of our analysis we set Newton's constant $ G=1 $.   

By solving the equations of motion, the Born-Infeld anti de sitter (BI AdS) solution may be found as,
\begin{equation}
ds^2 = -\chi dt^2+\chi ^{-1}dr^2+r^2 d\Omega^{2}\label{metric}
\end{equation} 
where,
\begin{eqnarray}
\chi(r) = 1-\frac{m}{r^{n-2}}+\left[\frac{4b^{2}}{n(n-1)} +\frac{1}{l^{2}}\right]r^{2} -\frac{2\sqrt{2}b}{n(n-1)r^{n-3}}\sqrt{2b^{2}r^{2n-2}+(n-1)(n-2)q^{2}}\nonumber\\
+\frac{2(n-1)q^{2}}{nr^{2n-4}}H\left[\frac{n-2}{2n-2},\frac{1}{2},\frac{3n-4}{2n-2},-\frac{(n-1)(n-2)q^{2}}{2b^{2}r^{2n-2}} \right] 
\label{chi},
\end{eqnarray}
and $ H $ is a hyper-geometric function \cite{hyper}. 

It is interesting to note that for $ n\geq 3 $ the metric (\ref{chi}) has a curvature singularity at $ r=0 $. It is in fact possible to show that this singularity is hidden behind the event horizon(s) whose location may be obtained through the condition $\chi(r) = 0 $. Therefore the above solution (\ref{metric}) could be interpreted as a space time with black holes \cite{dey}.

In the limit $ b\rightarrow\infty $ and $ Q\neq0 $ one obtains the corresponding solution for Reissner Nordstrom (RN) AdS black holes. Clearly this is a nonlinear generalization of the RN AdS black holes. Here $ m $ is related to the ADM mass ($ M $) of the black hole as \cite{cai},\cite{dey}, 
\begin{eqnarray}
M=\frac{(n-1)}{16\pi}\omega_{n-1}m\nonumber\\
\label{mq}
\end{eqnarray}
where $ \omega_{n-1}\left(=\frac{2\pi^{n/2}}{\Gamma(n/2)}\right) $ is the volume of the unit $(n-1)$ sphere. Identical expression could also be found for the RN AdS case \cite{rn3}.

Electric charge ($ Q $) may defined as \cite{cai},
\begin{equation}
Q=\frac{1}{4\pi}\int \ast F d\Omega 
\end{equation}
which finally yields,
\begin{equation}
Q=\frac{\sqrt{(n-1)(n-2)}}{4\pi\sqrt{2}}\omega_{n-1}q.
\end{equation}

 Using (\ref{mq}) one can rewrite (\ref{chi}) as,
\begin{eqnarray}
\chi(r) = 1-\frac{16\pi M}{(n-1)\omega_{n-1}r^{n-2}}+\frac{r^{2}}{l^{2}}+\frac{4b^{2}r^{2}}{n(n-1)}\left[ 1-\sqrt{1+\frac{16\pi^{2}Q^{2}}{b^{2}r^{2(n-1)}\omega_{n-1}^{2}}}\right] \nonumber\\
+\frac{64\pi^{2}Q^{2}}{n(n-2)r^{2n-4}\omega_{n-1}^{2}}H\left[\frac{n-2}{2n-2},\frac{1}{2},\frac{3n-4}{2n-2},-\frac{16\pi^{2}Q^{2}}{b^{2}r^{2n-2}\omega_{n-1}^{2}} \right] 
\label{chi1}.
\end{eqnarray}

In order to obtain an expression for the ADM mass ($ M $) of the black hole we set $\chi(r_{+})=0$, which yields, $ (G=1) $
\begin{eqnarray}
M= \frac{(n-1)\omega_{n-1}r_{+}^{n-2}}{16\pi}+\frac{\omega_{n-1}(n-1)r_{+}^{n}}{16\pi l^{2}} + \frac{b^{2}r_{+}^{n}\omega_{n-1}}{4\pi n}\left[ 1-\sqrt{1+\frac{16\pi^{2}Q^{2}}{b^{2}r_{+}^{2(n-1)}\omega_{n-1}^{2}}}\right]\nonumber\\
+\frac{4\pi Q^{2}(n-1)}{n(n-2)\omega_{n-1}r_{+}^{n-2}}\left[ 1-\frac{8\pi^{2}Q^{2}(n-2)}{b^{2}(3n-4)r_{+}^{2n-2}\omega_{n-1}^{2}}\right]+ O(1/b^{4}) 
\label{M}
\end{eqnarray}
where $ r_{+} $ is the radius of the outer event horizon. Here the parameter $ b $ is chosen in such a way so that $ \frac{1}{b^{2}}\ll 1 $ or, in other words our analysis is carried out in the large $ b $ limit. Also, such a limit is necessary for abstracting the results for RN AdS black holes obtained by taking $ b\rightarrow\infty $. Therefore all the higher order terms from $ (1/b^{2})^{2} $ onwards have been dropped out from the series expansion of $ H\left[\frac{n-2}{2n-2},\frac{1}{2},\frac{3n-4}{2n-2},-\frac{16\pi^{2}Q^{2}}{b^{2}r^{2n-2}\omega_{n-1}^{2}} \right] $. Moreover our results are valid upto an order $ (1/b^{2}) $. Using (\ref{mq}) one can express the electrostatic potential difference ($ \Phi $) between the horizon and infinity as,
\begin{eqnarray}
\Phi &=&\sqrt{\frac{n-1}{2n-4}}\frac{q}{r_{+}^{n-2}}H\left[\frac{n-2}{2n-2},\frac{1}{2},\frac{3n-4}{2n-2},-\frac{(n-1)(n-2)q^{2}}{2b^{2}r^{2n-2}} \right] \nonumber\\
&=&  \frac{4\pi Q}{(n-2)\omega_{n-1}r_{+}^{n-2}}\left[ 1-\frac{8\pi^{2}Q^{2}(n-2)}{b^{2}(3n-4)r_{+}^{2n-2}\omega_{n-1}^{2}}\right]+ O(1/b^{4})
\label{Phi}
\end{eqnarray}
where $ Q $ is the electric charge. Henceforth all our results are valid upto $ O(1/b^{2}) $ only.

Using (\ref{chi1}) and (\ref{M}), the Hawking temperature may be obtained as,
\begin{eqnarray}
T&=& \frac{\chi^{'}(r_{+})}{4\pi}\nonumber\\
&=&\frac{1}{4\pi}\left[ \frac{n-2}{r_{+}}+\frac{nr_{+}}{l^{2}}+\frac{4b^{2}r_{+}}{n-1}\left( 1-\sqrt{1+\frac{16\pi^{2}Q^{2}}{b^{2}r_+^{2(n-1)}\omega_{n-1}^{2}}}\right)   \right] \label{T}.
\end{eqnarray}
In the appropriate limit ($ b\rightarrow\infty $) the corresponding expression of Hawking temperature for the RN AdS black hole may be obtained as,
\begin{eqnarray}
T_{RN AdS}=\frac{1}{4\pi}\left[ \frac{n-2}{r_{+}}+\frac{nr_{+}}{l^{2}}-\frac{32\pi^{2}Q^{2}}{(n-1)r_{+}^{2n-3}\omega_{n-1}^{2}} \right] \label{Trn}.
\end{eqnarray}

Using (\ref{M}) and (\ref{T}) the entropy of the BI AdS black hole may be found as,
\begin{eqnarray}
S=\int T^{-1}\left( \frac{\partial M}{\partial r_{+}}\right)_{Q} dr_{+}= \frac{\omega_{n-1}r_{+}^{n-1}}{4}\label{S}. 
\end{eqnarray}
It is interesting to note that identical expression could also be found for the RN AdS case \cite{rn3}. 

In order to investigate the critical phenomena, it is necessary to compute the heat capacity at constant charge ($ C_{Q} $). Using (\ref{T}) and (\ref{S}) the specific heat (at constant charge) may be found as,
\begin{eqnarray}
C_{Q}=T\left(\frac{\partial S}{\partial T} \right)_{Q} = T \frac{\left( \partial S/\partial r_{+}\right)_{Q}}{\left( \partial T/\partial r_{+}\right)_{Q}}=\frac{\Im(r_+,Q)}{\Re(r_+,Q)}\label{CQ}
\end{eqnarray}
where,
\begin{eqnarray}
\Im(r_+,Q)=\frac{(n-1)\omega_{n-1}r_+^{3n-7}}{4}\sqrt{1+\frac{16\pi^{2}Q^{2}}{b^{2}r_+^{2(n-1)}\omega_{n-1}^{2}}}\nonumber\\
\times \left[(n-2)r_{+}^{2}+\frac{nr_{+}^{4}}{l^{2}}+
\frac{4b^{2}r_{+}^{4}}{n-1}\left( 1-\sqrt{1+\frac{16\pi^{2}Q^{2}}{b^{2}r_+^{2(n-1)}\omega_{n-1}^{2}}}\right)  \right] 
\end{eqnarray}
and,
\begin{eqnarray}
\Re(r_+,Q)= r_+^{2n-4}\left(\frac{nr_+^{2}}{l^{2}}-n+2 \right) \sqrt{1+\frac{16\pi^{2}Q^{2}}{b^{2}r_+^{2(n-1)}\omega_{n-1}^{2}}}+\left(\frac{n-2}{n-1} \right) \frac{64 \pi^{2}Q^{2}}{\omega_{n-1}^{2}} \nonumber\\
-\frac{4b^{2}r_{+}^{2n-2}}{n-1}\left( 1-\sqrt{1+\frac{16\pi^{2}Q^{2}}{b^{2}r_+^{2(n-1)}\omega_{n-1}^{2}}}\right)\label{d}.
\end{eqnarray}

From (\ref{CQ}) we note that in order to have a divergence in $ C_Q $ one must satisfy the following condition, 
\begin{eqnarray}
\Re(r_+,Q)= r_+^{2n-4}\left(\frac{nr_+^{2}}{l^{2}}-n+2 \right) \sqrt{1+\frac{16\pi^{2}Q^{2}}{b^{2}r_+^{2(n-1)}\omega_{n-1}^{2}}}+\left(\frac{n-2}{n-1} \right) \frac{64 \pi^{2}Q^{2}}{\omega_{n-1}^{2}} \nonumber\\
-\frac{4b^{2}r_{+}^{2n-2}}{n-1}\left( 1-\sqrt{1+\frac{16\pi^{2}Q^{2}}{b^{2}r_+^{2(n-1)}\omega_{n-1}^{2}}}\right)=0.\label{root}
\end{eqnarray}

Although it is quite difficult to solve (\ref{root}) analytically, one can attempt to solve this equation numerically. In order to do that it is first necessary to fix the parameters ($ b $ and $ Q $) of the theory. It is generally observed that the choice of parameters is not completely arbitrary in $ (3+1) $ dimension  \cite{myung}. Nevertheless one can give certain plausibility arguments that give a bound on the parameter space in order to have real positive roots for (\ref{root}). The boundedness of the parameter space could be achieved by demanding that a smooth extremal limit holds. In other words, we choose the parameters in such a way so that an extremal black hole could be found in the appropriate limit. Furthermore, once the choice of parameters has been determined in this manner, it is easy to show that meaningful results are not obtained in the non extremal case if one is not confined to this choice. Let us consider the extremal BI AdS black holes. Here both $ \chi(r) $ and $ \frac{d\chi}{dr} $ vanish at the degenerate horizon ($ r_e $). From the above two conditions and using (\ref{M}) we arrive at the following equation,  
\begin{equation}
1+\left(\frac{4b^{2}}{n-1} +\frac{n}{l^{2}}\right)\frac{r_{e}^{2}}{n-2}-\frac{4b^{2}r_{e}^{2}}{(n-1)(n-2)}\sqrt{1+\frac{16\pi^{2}Q^{2}}{b^{2}r_e^{2(n-1)}\omega_{n-1}^{2}}} =0.\label{extm}
\end{equation}   

For $ n=3 $, equation (\ref{extm}) can be solved analytically for $ r_{e}^{2} $ which results in the following bound on the parameter space of the BI AdS black hole as \cite{myung}, \cite{dibakar4},
\begin{equation}
0.5\leq bQ<\infty \label{bq}.
\end{equation}
For $ bQ<0.5 $, the root $ r_e^{2} $ becomes negative and hence there exists no real positive solution for $ r_e $ \cite{myung},\cite{dibakar4}. A detailed analysis of the critical phenomena for the non extremal case was done by us \cite{dibakar4} subject to the condition (\ref{bq}) that ensured a smooth extremal limit. We adopt a similar stance for higher dimensional black holes also. \\ \\ \\

\textbf{\textit{The $ n=4 $ case }}

For $ n=4 $, on the other hand, equation (\ref{extm}) turns out to be a cubic equation in the variable $ r_{e}^{2} $, which is indeed quite difficult to solve analytically. However, it is possible to solve equation (\ref{extm}) numerically for  $ r_{e}^{2} $. The solutions are provided below in a tabular form (table 1) for various choice of parameters ($ Q $ and $ b $).
\begin{table}[htb]
\caption{Numerical solutions for $r_e^{2}$ for $ n=4 $ and $ l=10 $ (extremal case)}   
\centering                          
\begin{tabular}{c c c c c c c}            
\hline\hline                        
$Q$ & $b$  & $r_{e1}^{2}$ &$r_{e2}^{2}$ & $r_{e3}^{2} $   \\ [0.05ex]
\hline
0.5 & 10 & -0.187184 & 0.180352 & 50.7835 \\
0.5 & 1 & -102.266 & -0.776247 & 0.04254 \\                              
0.5 & 0.5 & -6.74452 & -4.46669 & 0.0112109 \\
0.05 & 0.05 & -0.542572-i 4.97467 & -0.542572+i 4.97467  & $1.12579\times 10^{-6}$ \\  [0.05ex]         
\hline                              
\end{tabular}\label{E1}  
\end{table}
From the roots of equation (\ref{extm}) it is quite evident that for $ n=4 $ there is as such no bound on the parameter space of BI AdS black holes as far as a smooth extremal limit is concerned. It is also interesting to note that for $ bQ\leq 0.5 $ we have only one real positive root for $ r_e $, whereas for $ bQ>0.5 $ we have two real positive roots for $ r_e $. We are now in a position to find out the roots of the equation (\ref{root}) numerically for $ n=4 $ considering various values of the parameters ($ b $ and $ Q $), which are given below in table 2. 
\begin{table}[htb]
\caption{Roots of the equation (\ref{root}) for $ n=4 $ and $ l=10 $(non extremal case)}   
\centering                          
\begin{tabular}{c c c c c c c c c c c}            
\hline\hline                        
$Q$ & $b$  & $r_1$ &$r_2$ & $r_3 $ &$r_4$&$ r_5 $& $ r_6 $ \\ [0.05ex]
\hline
0.5&10&0.6410&7.0708&-7.0708&-0.6410&$5.9376\times 10^{-17}-i 0.641$&$5.9376\times 10^{-17}+i 0.641$\\ 
0.5&1&0.4538&7.0708&-7.0708&-0.4538&$-3.6375\times 10^{-17}-i 0.735$&$-3.6375\times10^{-17}+ i0.735$\\
0.5&0.5&0.2139&7.0708&-7.0708&-0.2139&$3.2485\times 10^{-17} -i 0.875$&$3.2485\times 10^{-17} +i 0.875$\\
0.05&0.05&0.0021&7.0710&-7.0710&-0.0021&$ 3.3787\times10^{-17} - i0.860$&$3.3787\times10^{-17} + i 0.860$\\
\\ [0.05ex]         
\hline                              
\end{tabular}
\label{E1}          
\end{table} 

Two crucial points are to be noted at this stage- (i) $ C_Q $ (\ref{CQ}) possesses only simple poles and (ii) there are always two real positive roots ($ r_1 $ and $ r_2 $) of (\ref{root}) for different choice of parameters. These two roots ($ r_1 $ and $ r_2 $) correspond to the critical points for the phase transition phenomena occurring in BI AdS black holes. For our detailed analysis we choose $ Q=0.5 $  and $ b=10 $, corresponding to the first row in table 2. With this particular choice the critical points are found to be  $ r_1= 0.6410$ and $ r_2=7.0708 $, which are also depicted in various figures (1 and 2).
\begin{figure}[h]
\centering
\includegraphics[angle=0,width=8cm,keepaspectratio]{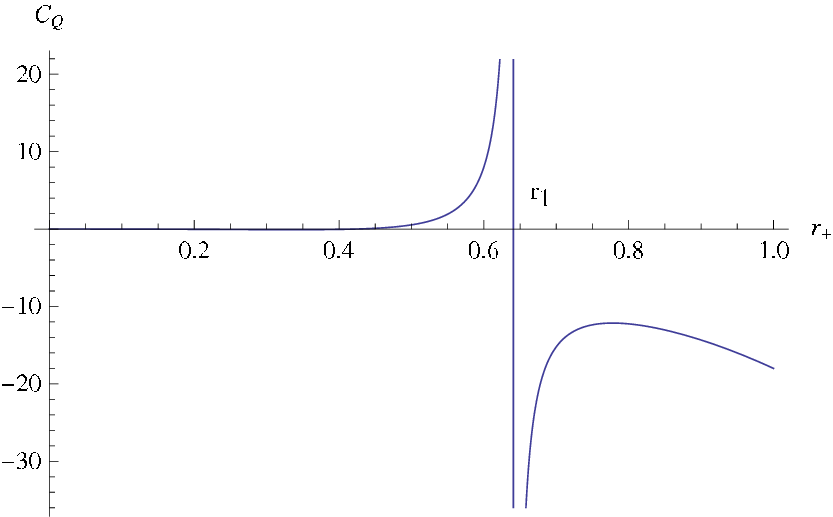}
\caption[]{\it Discontinuity of Specific heat ($ C_{Q} $) for Born-Infeld AdS black hole at $r_{+}= r_1$ for $Q(=Q_c)=0.5$, $ b=10 $, $ n=4 $ and $ l=10 $}
\label{figure 2a}
\end{figure}
\begin{figure}[h]
\centering
\includegraphics[angle=0,width=8cm,keepaspectratio]{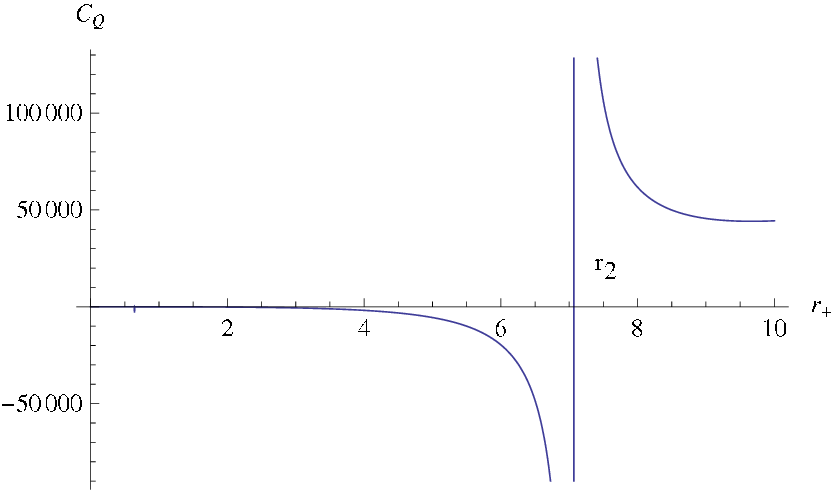}
\caption[]{\it Discontinuity of Specific heat ($ C_{Q} $) for Born-Infeld AdS black hole at $r_{+}= r_2$ for $Q(=Q_c)=0.5$, $ b=10 $, $ n=4 $ and $ l=10 $}
\label{figure 2a}
\end{figure}

From figures 1 and 2 we observe that $ C_{Q} $ suffers discontinuities exactly at two points, namely $ r_{1} $ and $ r_{2} $(discussed in the previous paragraph), which may be identified as the critical points for the phase transition phenomena in BI AdS black holes. From these figures we note that there is a sign flip in the heat capacity around $ r_i(i=1,2) $, which indicates the onset of a continuous higher order transition near critical points. Using a grand canonical framework, one can in fact show that BI AdS black holes undergo a second order phase transition near the critical point \cite{dibakar3}.  From figures 1 and 2 we note that $ C_Q $ is positive for $ r>r_2 $ and $ r<r_1 $, while it is negative for $ r_1<r<r_2 $. It is the positive heat capacity that corresponds to a thermodynamically stable phase, whereas, on the other hand, negative heat capacity stands for a thermodynamically unstable phase. Also, since the black hole with larger mass possesses a larger horizon area/radius, therefore $ r_1 $ corresponds to the critical point for the transition between a lower mass (stable) black hole to an intermediate higher mass (unstable) black hole. On the other hand $ r_2 $ stands for the critical point for the transition between the intermediate unstable black hole to a higher mass (stable) black hole. 

Similar features may also be observed for the RN AdS case. The expression for the heat capacity is obtained by taking the $ b\rightarrow\infty $ limit of (\ref{CQ}). One obtains,
\begin{equation}
\left(C_Q \right)_{RN AdS}=\frac{(n-1)\omega_{n-1}r_+^{3n-7}\left[(n-2)r_{+}^{2}+\frac{nr_{+}^{4}}{l^{2}}-\frac{32\pi^{2}Q^{2}}{(n-1)r_{+}^{2n-6}\omega_{n-1}^{2}}\right]}{4\left[ r_+^{2n-4}\left(\frac{nr_+^{2}}{l^{2}}-n+2 \right)+\frac{32(2n-3) \pi^{2}Q^{2}}{(n-1)\omega_{n-1}^{2}}\right]}. 
\end{equation}
The plot of $ C_Q $ vs $ r_+ $ are given in figures 3 and 4,
\begin{figure}[h]
\centering
\includegraphics[angle=0,width=8cm,keepaspectratio]{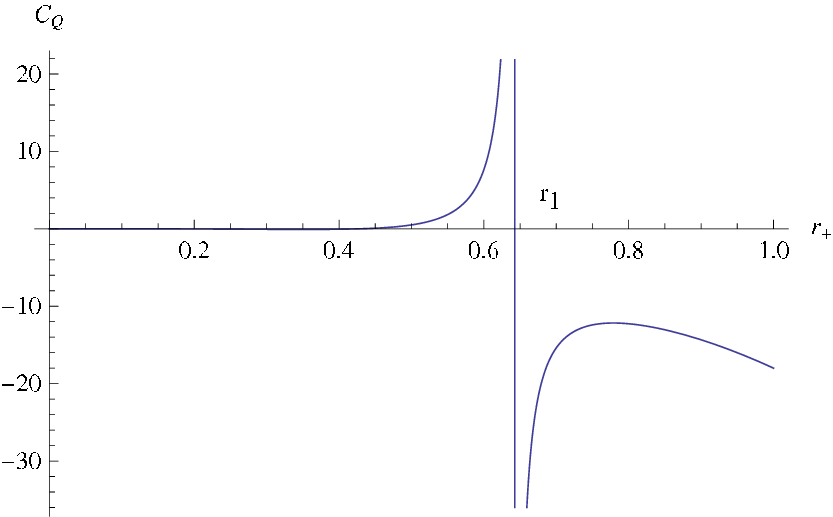}
\caption[]{\it Discontinuity of Specific heat ($ C_{Q} $) for Reissner Nordstrom AdS black hole at $r_{+}= r_1$ for $Q(=Q_c)=0.5$, $ n=4 $ and $ l=10 $}
\label{figure 2a}
\end{figure}
\begin{figure}[h]
\centering
\includegraphics[angle=0,width=8cm,keepaspectratio]{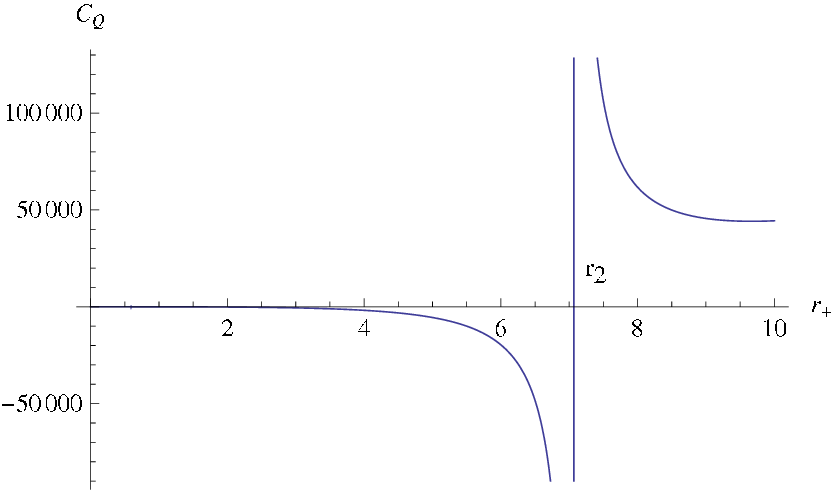}
\caption[]{\it Discontinuity of Specific heat ($ C_{Q} $) for Reissner Nordstrom AdS black hole at $r_{+}= r_2$ for $Q(=Q_c)=0.5$, $ n=4 $ and $ l=10 $}
\label{figure 2a}
\end{figure}
The nature of the phase transition is similar to BI AdS case.\\ 

\textbf{\textit{The $ n=5 $ case }}

For $ n=5 $, equation (\ref{extm}) turns out to be a quartic equation in the variable $ r_e^{2} $. The corresponding roots  ($ r_{e}^{2} $) of the equation (\ref{extm}) are provided in the following tabular form (table 3). From table 3 we note that, even for $ bQ<0.5 $ one can have two real positive roots of the equation (\ref{extm}) in $ (5+1) $ dimensions. However one should note that for $ bQ\rightarrow0 $ the number of real positive roots again reduces to one (see, for instance, the last row in table 3). 
\begin{table}[htb]
\caption{Numerical solutions for $r_e^{2}$ for $ n=5 $ and $ l=10 $(extremal case)}   
\centering                          
\begin{tabular}{c c c c c c c c}            
\hline\hline                        
$Q$ & $b$  & $r_{e1}^{2}$ &$r_{e2}^{2}$ & $r_{e3}^{2} $ & $ r_{e4}^{2} $   \\ [0.05ex]
\hline
0.5 & 10 & -0.11104-0.183039 i & -0.11104+0.183039 i & 0.207093 & 60.06 \\
0.5 & 1 & -1.39385 & -0.0819668 & 0.0775052 & 66.0137 \\                              
0.5 & 0.5 & -4.72104 & -0.0399487 & 0.0396321 & 84.7214 \\
0.05 & 0.05 & -100 & -40  & -0.00039789 & 0.000397885 \\  [0.05ex]         
\hline                              
\end{tabular}\label{E1}  
\end{table}

Finally, we aim to find out the roots of the equation (\ref{root}) for different choice of parameters ($ Q $ and $ b $), which essentially gives us the critical points for the phase transition in the non extremal regime.
\begin{table}[htb]
\caption{Roots of the equation (\ref{root}) for $ n=5 $ and $ l=10 $(non extremal case)}   
\centering                          
\begin{tabular}{c c c c c c c c c c c}            
\hline\hline                        
$Q$ & $b$  & $r_1$ &$r_2$ & $r_3 $ &$r_4$&$ r_{5,6} $& $ r_{7,8} $ \\ [0.005ex]
\hline
0.5&10&0.63&7.74&-7.74&-0.63&$-0.31\pm i 0.55$ &$0.31\pm i 0.55$  \\ 
0.5&1&0.49&7.74&-7.74&-0.49&$-0.29\pm i 0.64$&$0.29\pm i 0.64$\\
0.5&0.5&0.34&7.74&-7.74&-0.34&$-0.32\pm i 0.76$&$0.32\pm i 0.76$\\
0.05&0.05&0.03&7.74&-7.74&-0.03&$ -0.31\pm i 0.76 $&$0.31\pm i 0.76 $\\
\\ [0.005ex]         
\hline                              
\end{tabular}
\label{E1}          
\end{table}

Like in the earlier case, from table 4 we observe that $ C_Q $(\ref{CQ}) possesses simple poles, two of which are real positive  ($ r_1 $ and $ r_2 $) that may be regarded as the critical points corresponding to the phase transition phenomena in (non extremal) BI AdS black holes. Taking $ Q=0.5 $ and $ b=10 $, these critical points have been shown explicitly in figures (5 and 6).
\begin{figure}[h]
\centering
\includegraphics[angle=0,width=8cm,keepaspectratio]{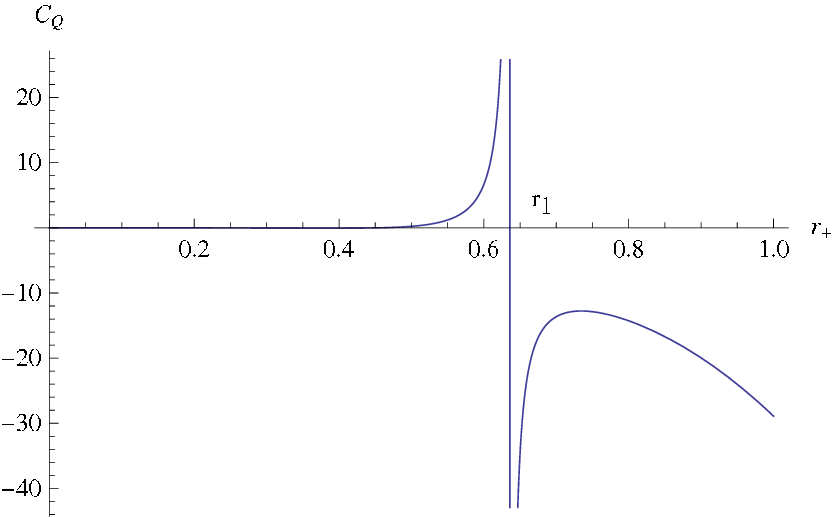}
\caption[]{\it Discontinuity of Specific heat ($ C_{Q} $) for Born-Infeld AdS black hole at $r_{+}= r_1$ for $Q(=Q_c)=0.5$, $ b=10 $, $ n=5 $ and $ l=10 $}
\label{figure 2a}
\end{figure}
\begin{figure}[h]
\centering
\includegraphics[angle=0,width=8cm,keepaspectratio]{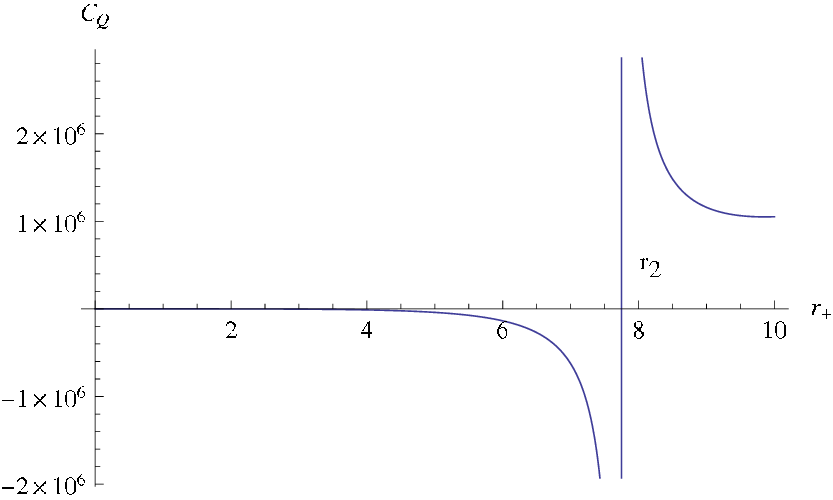}
\caption[]{\it Discontinuity of Specific heat ($ C_{Q} $) for Born-Infeld AdS black hole at $r_{+}= r_2$ for $Q(=Q_c)=0.5$, $ b=10 $, $ n=5 $ and $ l=10 $}
\label{figure 2a}
\end{figure}

Likewise, in the appropriate limit ($ b\rightarrow\infty $), one can also obtain the corresponding critical points for the RN AdS case which are shown in various figures (7 and 8).
\begin{figure}[h]
\centering
\includegraphics[angle=0,width=8cm,keepaspectratio]{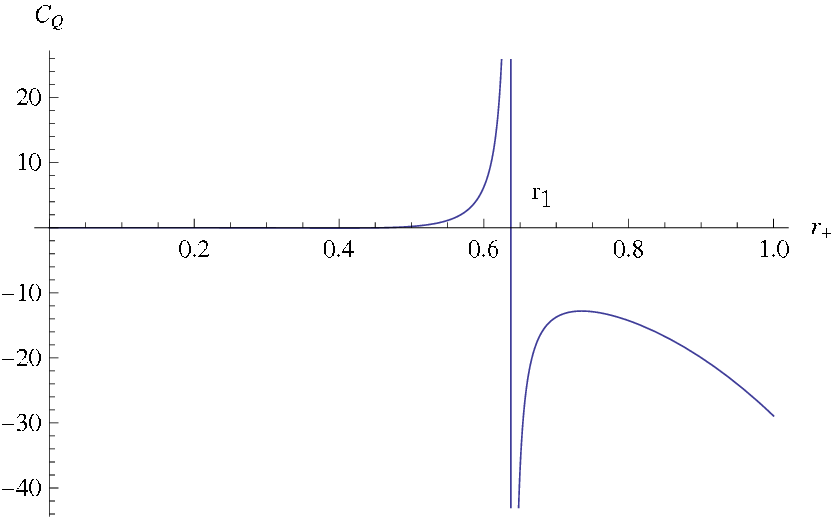}
\caption[]{\it Discontinuity of Specific heat ($ C_{Q} $) for Reissner Nordstrom AdS black hole at $r_{+}= r_1$ for $Q(=Q_c)=0.5$, $ n=5 $ and $ l=10 $}
\label{figure 2a}
\end{figure}
\begin{figure}[h]
\centering
\includegraphics[angle=0,width=8cm,keepaspectratio]{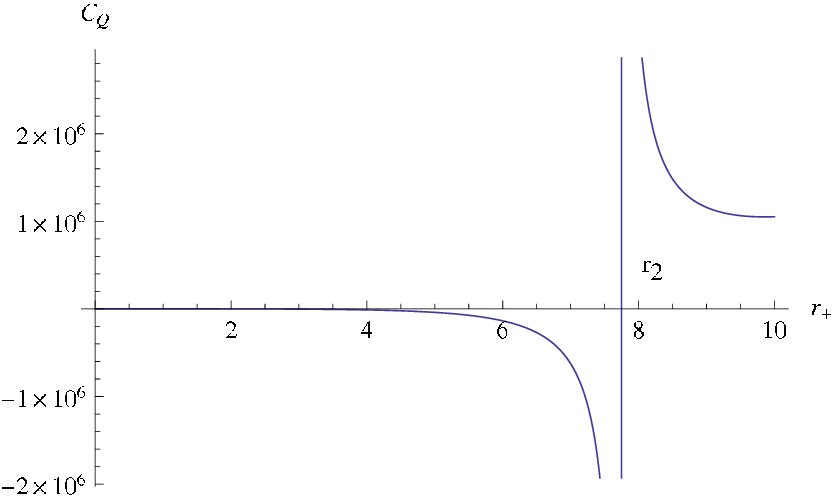}
\caption[]{\it Discontinuity of Specific heat ($ C_{Q} $) for Reissner Nordstrom AdS black hole at $r_{+}= r_2$ for $Q(=Q_c)=0.5$, $ n=5 $ and $ l=10 $}
\label{figure 2a}
\end{figure}

Therefore, from the above analysis it is quite suggestive that the boundedness on the parameter space, which exists in $ (3+1) $ dimension, eventually disappears in higher dimensions which is also consistent with our finding of the critical points in the corresponding non extremal case for various choice of parameters.

\section{Critical exponents and scaling laws in higher dimensions}
In the usual theory of phase transitions it is customary to study the behavior of a given system close to its critical point by means of a set of critical exponents $ (\alpha,\beta,\gamma,\delta,\varphi,\psi,\nu,\eta) $. These critical exponents determine the qualitative nature of the critical behavior of the given system in the neighborhood of the critical point. By virtue of the so called scaling laws, one can in fact see that only two of the eight critical exponents are actually independent. Based on the renormalization group approach, one can calculate these critical exponents. Systems having identical critical exponents have similar critical behavior and hence fall into the same universality class. Here we show similar conclusions may be drawn for black holes also. We find that, apart from $ \delta $ that takes the value 2, all other critical exponents are $ \frac{1}{2} $. The result is independent of the dimensionality of space time. This shows that BI AdS black holes (and hence RN AdS black holes) in arbitrary dimensions fall in the same universality class.

 In order to calculate the critical exponent ($ \alpha $) that is associated with the divergences of the heat capacity ($ C_{Q} $), we first note that near the critical points ($ r_i $) we can write
\begin{equation}
r_{+}=r_{i}(1+\Delta), ~~~~ i=1,2 \label{tri}
\end{equation}
where $|\Delta| <<1 $.
\begin{figure}[h]
\centering
\includegraphics[angle=0,width=7cm,keepaspectratio]{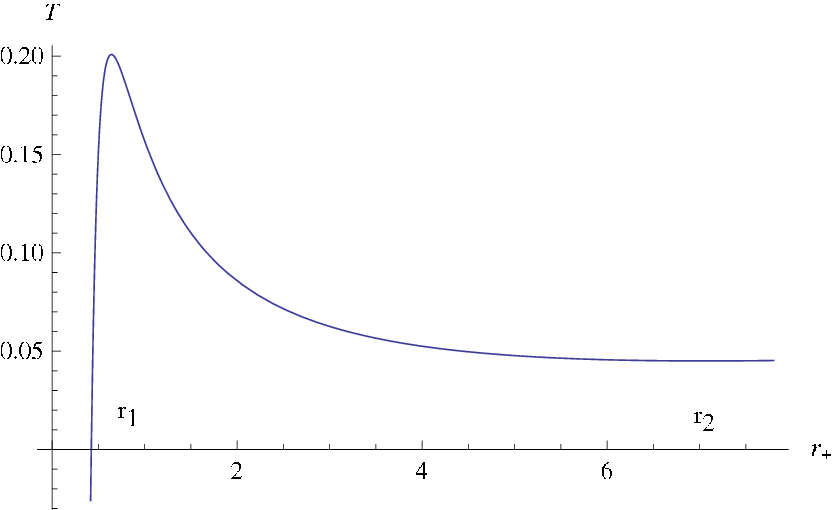}
\caption[]{\it Temperature ($T$) plot for Born-Infeld AdS black hole for $Q(=Q_c)=0.5$, $ b=10 $, $ n=4 $ and $ l=10 $}
\label{figure 2a}
\end{figure}
\begin{figure}[h]
\centering
\includegraphics[angle=0,width=7cm,keepaspectratio]{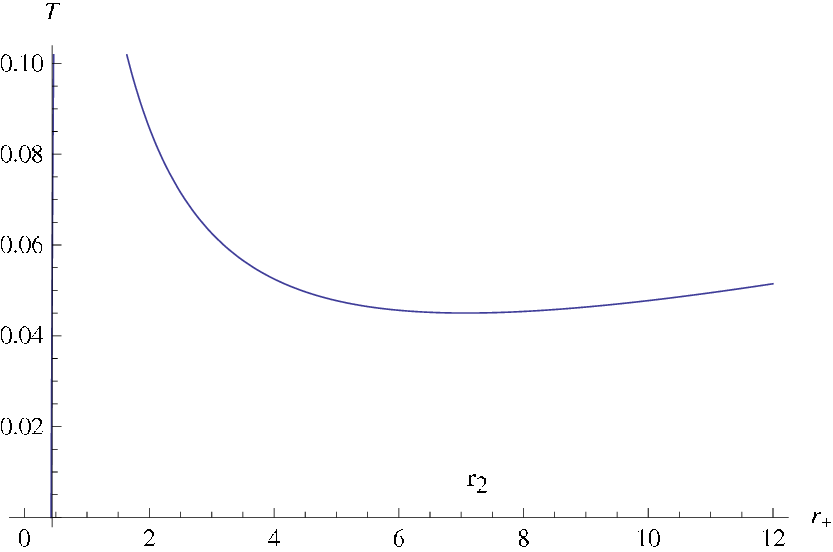}
\caption[]{\it Change in slope in the temperature ($T$) plot at $ r_+=r_2 $ for Born-Infeld AdS black hole for $Q(=Q_c)=0.5$, $ b=10 $, $ n=4 $ and $ l=10 $}
\label{figure 2a}
\end{figure}
\begin{figure}[h]
\centering
\includegraphics[angle=0,width=7cm,keepaspectratio]{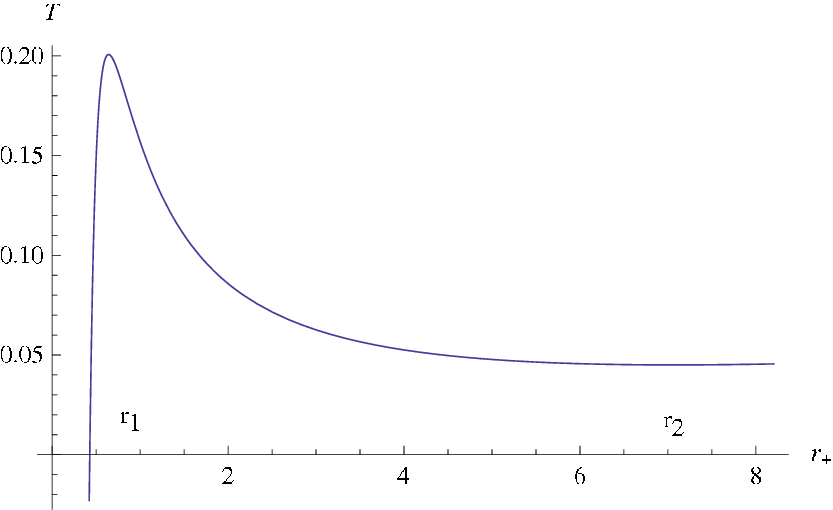}
\caption[]{\it Temperature ($T$) plot for Reissner Nordstrom AdS black hole for $Q(=Q_c)=0.5$, $ n=4 $ and $ l=10 $}
\label{figure 2a}
\end{figure}
\begin{figure}[h]
\centering
\includegraphics[angle=0,width=7cm,keepaspectratio]{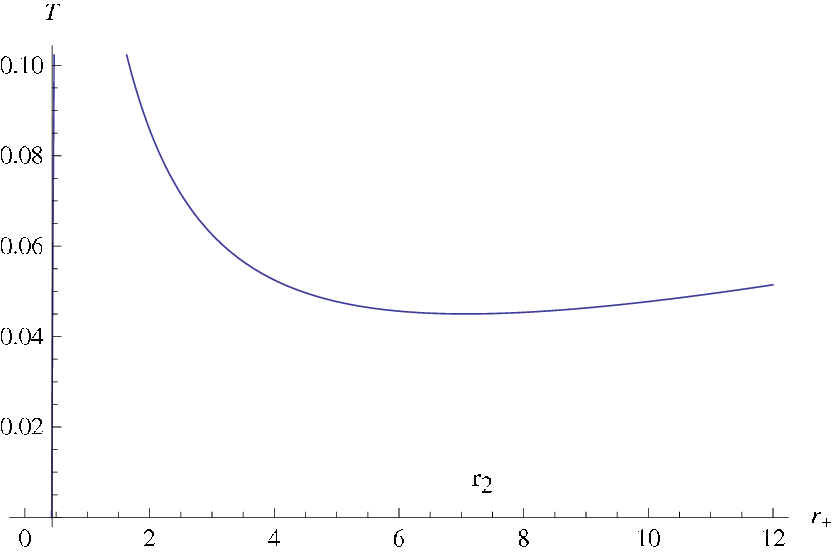}
\caption[]{\it Change in slope in the temperature ($T$) plot at $ r_+=r_2 $ for Reissner Nordstrom AdS black hole for $Q(=Q_c)=0.5$, $ n=4 $ and $ l=10 $}
\label{figure 2a}
\end{figure}
As already discussed there are two distinct positive roots for the critical point $ r_i $ ($ r_1 $ and $ r_2 $). Also, any function of $ r_+ $, in particular the temperature $ T(r_+) $, may be expressed as
\begin{equation}
T(r_+)=T(r_i)(1+\epsilon)
\end{equation}
where $ |\epsilon|<<1$.
As a next step, for a fixed value of the charge ($ Q $), we Taylor expand $ T(r_+) $ in a sufficiently small neighborhood of $ r_i $ which yields,
\begin{eqnarray}
T(r_+)= T(r_i)+\left[ \left( \frac{\partial T}{\partial r_+}\right)_{Q=Q_c}\right]_{r_+=r_i} (r_+-r_i)+\frac{1}{2} \left[ \left( \frac{\partial^{2} T}{\partial r^{2}_+}\right)_{Q=Q_c}\right]_{r_+=r_i} (r_+-r_i)^{2}\nonumber\\
+ higher~~ order~~ terms.\label{Tr}
\end{eqnarray}
Since $ C_Q $ diverges at $ r_+=r_i $, therefore the second term on the R.H.S. of (\ref{Tr}) vanishes by virtue of equation (\ref{CQ}). This fact has also been depicted in various other figures (see figures 5 and 6).  
Using (\ref{tri}) we finally obtain from (\ref{Tr})
\begin{equation}
\Delta=\frac{\epsilon^{1/2}}{D_{i}^{1/2}}\label{delta}
\end{equation} 
where\footnote{We use the notation $ T(r_i)=T_i $.},
\begin{eqnarray}
D_i =\frac{r^{2}_{i}}{2T_{i}}\left[ \left( \frac{\partial^{2} T}{\partial r^{2}_+}\right)_{Q=Q_c}\right]_{r_+=r_i}=\frac{\Sigma(r_i,Q_c)}{4\pi r^{2n-3}_{i}T_i \left(1+\frac{16 \pi^{2} Q^{2}_{c}}{{b^{2}r^{2n-2}_{i}\omega_{n-1}^{2}}}\right)^{3/2}}
\end{eqnarray}
with,
\begin{eqnarray}
\Sigma(r_i,Q_c) = (n-2)r^{2n-4}_{i}\left(1+\frac{16\pi^{2} Q^{2}_{c}}{{b^{2}r^{2n-2}_{i}\omega_{n-1}^{2}}}\right)^{3/2}+\frac{512(n-1)\pi^{4}Q^{4}_{c}}{b^{2}r^{2n-2}_{i}\omega_{n-1}^{4}}\nonumber\\
-32(2n-3)\pi^{2}Q_{c}^{2}\omega_{n-1}^{-2}\left(1+\frac{16\pi^{2} Q^{2}_{c}}{b^{2}r^{2n-2}_{i}\omega_{n-1}^{2}}\right).
\end{eqnarray}
In the limit $ b\rightarrow\infty $, the corresponding expression for RN AdS case may be obtained as,
\begin{eqnarray}
\left[D_i \right]_{RN AdS}=\frac{(n-2)r^{2n-4}_{i}-32(2n-3)\pi^{2}Q_{c}^{2}\omega_{n-1}^{-2}}{4\pi r^{2n-3}_{i}[T_i]_{RN AdS}}. 
\end{eqnarray}

A closer look at figures 5 and 6 reveals that in the neighborhood of $ r_+=r_2 $ we always have $ T(r_+)>T(r_2) $ so that $ \epsilon $ is positive. On the other hand for any point close to  $ r_+=r_1 $ we have $ T(r_+)<T(r_1) $ implying that $ \epsilon $ is negative. We will exploit these observations to find $ C_Q $ near the critical points.

Let us first compute the value of $ C_Q $ near the critical point $ r_+=r_2 $ (where $ \epsilon $ is positive). Substituting $ r_+ $ from (\ref{tri}) into (\ref{CQ}) we obtain,
\begin{equation}
C_Q=\frac{\Im(r_i(1+\Delta),Q_c)}{\Re(r_i(1+\Delta),Q_c)}.
\end{equation}
Taylor expanding around the critical point $\Delta\approx 0$ we obtain,
\begin{eqnarray}
C_Q \simeq ~~~\left[ \frac{A_{i}}{\epsilon^{1/2}}\right] _{r_i=r_2}
\end{eqnarray}
where,
\begin{equation}
A_i=\frac{D^{1/2}_{i} \zeta(r_i,Q_c)}{\xi(r_i,Q_c)}
\end{equation}
with,
\begin{eqnarray}
\zeta(r_i,Q_c)= \frac{(n-1)\omega_{n-1}r_{i}^{3n-7}}{8}\sqrt{1+\frac{16\pi^{2}Q_{c}^{2}}{b^{2}r_{i}^{2n-2}\omega_{n-1}^{2}}}\nonumber\\
\times\left[(n-2)r_{i}^{2}+\frac{nr_{i}^{4}}{l^{2}}+\frac{4b^{2}r_{i}^{4}}{n-1}\left( 1-\sqrt{1+\frac{16\pi^{2}Q_{c}^{2}}{b^{2}r_{i}^{2n-2}\omega_{n-1}^{2}}}\right)  \right] 
\end{eqnarray}
and,

\begin{eqnarray}
\xi(r_i,Q_c)= r_{i}^{2n-4}\left[(n-1)\frac{nr_{i}^{2}}{l^{2}}-(n-2)^{2} \right]\sqrt{1+\frac{16\pi^{2}Q_{c}^{2}}{b^{2}r_{i}^{2n-2}\omega_{n-1}^{2}}}\nonumber\\
-\frac{8(n-1)\pi^{2}Q_{c}^{2}}{b^{2}r_{i}^{2}\omega_{n-1}^{2}}\left(2-n+\frac{nr_{i}^{2}}{l^{2}} \right).  
\end{eqnarray}
 On the other hand, following a similar approach, the singular behavior of $ C_Q $ near $ r_+=r_1 $ (where $ \epsilon $ is negative) may be expressed as,
\begin{equation}
C_Q \simeq  ~~~\left[ \frac{A_{i}}{(-\epsilon)^{1/2}}\right] _{r_i=r_1}.
\end{equation}
Combining both of these facts into a single expression, we may therefore express the singular behavior of the heat capacity ($ C_Q $) near the critical points as,
\begin{eqnarray}
C_Q &\simeq & ~~~ \frac{A_{i}}{|\epsilon|^{1/2}}\nonumber\\
&=&~~~\frac{A_{i}T^{1/2}_{i}}{|T-T_i|^{1/2}}\label{CQ1}.
\end{eqnarray}

Similar arguments also hold for the RN AdS case (see figures 7 and 8). In order to obtain the corresponding expression for the singular behavior of the heat capacity near the critical points, we set $ b\rightarrow\infty $, which finally yields,
\begin{equation}
\left[C_Q \right]_{RN AdS}\simeq \frac{[A_{i}]_{RN AdS}[T^{1/2}_{i}]_{RN AdS}}{|T-T_i|^{1/2}}\label{rnads1}
\end{equation}
where,
\begin{eqnarray}
[A_{i}]_{RN AdS}=[D_i^{1/2}]_{RN AdS}\frac{(n-1)\omega_{n-1}r_{i}^{n-3}\left[(n-2)r_{i}^{2}+\frac{nr_{i}^{4}}{l^{2}}-\frac{32\pi^{2}Q_{c}^{2}}{(n-1)r_{i}^{2n-6}\omega_{n-1}^{2}}\right]}{8\left[(n-1)\frac{nr_{i}^{2}}{l^{2}}-(n-2)^{2} \right]}.
\end{eqnarray}

Comparing (\ref{CQ1}) and (\ref{rnads1}) with the standard form
\begin{equation}
C_Q \sim |T-T_i|^{-\alpha}
\end{equation}
we find $ \alpha=1/2 $.

Next, we want to calculate the critical exponent $ \beta $ which is related to the electric potential ($ \Phi $) for a fixed value of charge as,
\begin{equation}
\Phi(r_+) - \Phi(r_i) \sim |T-T_i|^{\beta}\label{Ph}.
\end{equation}
In order to do that we Taylor expand $ \Phi(r_+) $ close to the critical point $ r_+=r_i $ which yields, 
\begin{eqnarray}
\Phi(r_+)= \Phi(r_i)+\left[ \left( \frac{\partial \Phi}{\partial r_+}\right)_{Q=Q_c}\right]_{r_+=r_i} (r_+-r_i)
+ higher~~ order~~ terms.\label{phir}
\end{eqnarray}
Ignoring all the higher order terms in (\ref{phir}) and using (\ref{Phi}) and (\ref{delta}) we finally obtain
\begin{equation}
\Phi(r_+)- \Phi(r_i)= - \left( \frac{4\pi Q_c}{r_i^{n-2}\omega_{n-1} T^{1/2}_i D^{1/2}_{i}}\right) \left(1-\frac{8\pi^{2}Q^{2}_{c}}{b^{2}r^{2n-2}_{i}\omega_{n-1}^{2}} \right) |T-T_i|^{1/2}\label{Pr}.
\end{equation}
For the RN AdS case ($ b\rightarrow\infty $) the corresponding expression becomes,
\begin{eqnarray}
\Phi(r_+)- \Phi(r_i)= -  \frac{4\pi Q_c}{r_i^{n-2}\omega_{n-1} [T^{1/2}_i]_{RN AdS} [D^{1/2}_{i}]_{RN AdS}}  |T-T_i|^{1/2}\label{rnads2}.
\end{eqnarray}  
Comparing (\ref{Pr}) and (\ref{rnads2}) with (\ref{Ph}) we find $ \beta=1/2 $.

We next calculate the critical exponent $ \gamma $ which is related to the singular behavior of the isothermal compressibility related derivative $ K^{-1}_{T} $ (near the critical points $ r_i $) for a fixed value of charge ($ Q=Q_c $). This is defined as,
\begin{equation}
K^{-1}_{T}\sim |T-T_i|^{-\gamma}\label{k}
\end{equation}
In order to calculate $ K^{-1}_{T} $ we first note that,
\begin{eqnarray}
 K^{-1}_{T}=Q\left( \partial\Phi /\partial Q\right)_{T} =- Q \left(\frac{\partial \Phi}{\partial T} \right)_{Q} \left(\frac{\partial T}{\partial Q} \right)_{\Phi},
\end{eqnarray}
where we have used the thermodynamic identity $\left(\frac{\partial \Phi}{\partial T} \right)_{Q}\left(\frac{\partial T}{\partial Q} \right)_{\Phi}\left(\frac{\partial Q}{\partial \Phi} \right)_{T} =-1$.

Finally, using (\ref{Phi}) and (\ref{T}) the expression for $ K^{-1}_{T} $ may be found as,
\begin{equation}
K^{-1}_{T}=\frac{4\pi Q}{(n-2)\omega_{n-1}r_{+}^{n-2}}\left( 1-\frac{24\pi^{2}Q^{2}(n-2)}{(3n-4)\omega_{n-1}^{2}b^{2}r_{+}^{2n-2}}\right) \frac{\wp(Q,r_+)}{\Re(Q,r_+)}\label{kt}
\end{equation}
where,
\begin{eqnarray}
\wp(Q,r_+)= r_{+}^{2n-2}\left(\frac{n}{l^{2}}-\frac{n-2}{r_{+}^{2}} \right) \sqrt{1+\frac{16\pi^{2}Q^{2}}{b^{2}r_+^{2(n-1)}\omega_{n-1}^{2}}}+ \frac{1024\pi^{4}Q^{4}(n-2)}{r_{+}^{2n-2}b^{2}\omega_{n-1}^{4}(n-1)(3n-4)}\nonumber\\
-\frac{4b^{2}r_{+}^{2n-2}}{n-1}\left( 1-\sqrt{1+\frac{16\pi^{2}Q^{2}}{b^{2}r_+^{2(n-1)}\omega_{n-1}^{2}}}\right).
\end{eqnarray}
Taking the appropriate limit ($ b\rightarrow\infty $) one can easily obtain the corresponding expression of $K^{-1}_{T}$ for the RN AdS black hole as,
\begin{equation}
\left( K^{-1}_{T}\right)_{RN AdS} =\frac{4\pi Q}{(n-2)\omega_{n-1}r_{+}^{n-2}}\frac{\left[r_{+}^{2n-2}\left(\frac{n}{l^{2}}-\frac{n-2}{r_{+}^{2}} \right)+\frac{32\pi^{2}Q^{2}}{(n-1)\omega_{n-1}^{2}} \right] }{\left[r_+^{2n-4}\left(\frac{nr_+^{2}}{l^{2}}-n+2 \right)+\frac{32(2n-3) \pi^{2}Q^{2}}{(n-1)\omega_{n-1}^{2}} \right] }.
\end{equation}
From the above expressions it is interesting to note that both $ K^{-1}_{T} $ and the heat capacity ($ C_{Q} $) posses common singularities. It is interesting to note that a similar feature may also be observed for the Kerr Newmann black hole in asymptotically flat space \cite{cr5}. It is reassuring to note that all these features are in general compatible with standard thermodynamic systems\cite{dt}.  

Following our previous approach, we substitute $ r_+ $ from (\ref{tri}) into (\ref{kt}) and use (\ref{delta}) which finally yields,
\begin{equation}
K^{-1}_{T}=\frac{B_i}{|\epsilon|^{1/2}}= \frac{B_i T^{1/2}_i}{|T-T_i|^{1/2}}\label{KT}
\end{equation}   
where,
\begin{equation}
B_i=\frac{2\pi Q_c}{(n-2)\omega_{n-1}r_{i}^{n-2}}\left( 1-\frac{24\pi^{2}Q_{c}^{2}(n-2)}{(3n-4)\omega_{n-1}^{2}b^{2}r_{i}^{2n-2}}\right)\frac{D^{1/2}_i\wp(Q_c,r_i)}{\xi(Q_c,r_i)}.
\end{equation}
In the appropriate limit ($ b\rightarrow\infty $) the corresponding expression for the RN AdS black hole may be obtained as,
\begin{equation}
[K^{-1}_{T}]_{RN AdS}=\frac{[B_i]_{RN AdS}}{|\epsilon|^{1/2}}= \frac{[B_i]_{RN AdS} [T^{1/2}_i]_{RN AdS}}{|T-T_i|^{1/2}}\label{rnads3}
\end{equation}
where,
\begin{equation}
[B_i]_{RN AdS}= \frac{2\pi Q_c[D_{i}^{1/2}]_{RN AdS}}{(n-2)\omega_{n-1}r_{i}^{3n-6}}\frac{\left[r_{i}^{2n-2}\left(\frac{n}{l^{2}}-\frac{n-2}{r_{i}^{2}} \right)+\frac{32\pi^{2}Q_{c}^{2}}{(n-1)\omega_{n-1}^{2}}\right]}{\left[(n-1)\frac{nr_{i}^{2}}{l^{2}}-(n-2)^{2} \right]}.
\end{equation}
Comparing (\ref{KT}) and (\ref{rnads3}) with (\ref{k}) we note that $ \gamma=1/2 $.

Let us now calculate the critical exponent ($ \delta $) which is related to the electric potential ($ \Phi $) for the fixed value of the temperature $ T=T_i $ as,
\begin{equation}
\Phi(r_+) - \Phi(r_i) \sim |Q-Q_i|^{1/\delta}\label{PQ},
\end{equation}
where $ Q_i $ is the value of charge ($ Q $) at $ r_+=r_i $. We expand $ Q(r_+) $ in a sufficiently small neighborhood of $r_+= r_i $ which yields,
 \begin{eqnarray}
Q(r_+)= Q(r_i)+\left[ \left( \frac{\partial Q}{\partial r_+}\right)_{T=T_i}\right]_{r_+=r_i} (r_+-r_i)+\frac{1}{2} \left[ \left( \frac{\partial^{2} Q}{\partial r^{2}_+}\right)_{T=T_i}\right]_{r_+=r_i} (r_+-r_i)^{2}\nonumber\\
+ higher~~ order~~ terms.\label{Q}
\end{eqnarray}
Using the functional form
\begin{equation}
T=T(r_+,Q)
\end{equation}
and following our previous argument we note that,
\begin{equation}
\left[ \left( \frac{\partial Q}{\partial r_+}\right)_{T}\right]_{r_+=r_i}=-\left[ \left( \frac{\partial T}{\partial r_+}\right)_{Q}\right]_{r_+=r_i} \left( \frac{\partial Q}{\partial T}\right)_{r_+=r_i}=0 \label{e2}.
\end{equation}
Also note that near the critical point we can express the charge ($ Q $) as,
\begin{equation}
Q(r_+)= Q(r_i)(1+\Pi) \label{Qr}
\end{equation}
with $ |\Pi|<<1 $.  
Finally using (\ref{tri}), (\ref{Q}) and (\ref{Qr}) we obtain 
\begin{equation}
\Delta=\left(\frac{2Q_i}{M_i r^{2}_{i}} \right)^{1/2}\Pi^{1/2}\label{e4}
\end{equation}  
where,
\begin{eqnarray}
M_i=\left[ \left( \frac{\partial^{2} Q}{\partial r^{2}_+}\right)_{T}\right]_{r_+=r_i}
=\frac{(n-1)\omega_{n-1}^{2}\Gamma(r_i,Q_c)}{64\pi^{2}r_iQ_c\sqrt{1+\frac{16\pi^{2}Q_{c}^{2}}{b^{2}r_{i}^{2n-2}\omega_{n-1}^{2}}}} \label{Mi}
\end{eqnarray}
with,
\begin{eqnarray}
\Gamma(r_i,Q_c)=r_{i}^{2n-4}\left(1+\frac{16\pi^{2}Q_{c}^{2}}{b^{2}r_{i}^{2n-2}\omega_{n-1}^{2}} \right)\left[ \frac{n}{l^{2}}(2n-3)r_{i}^{2}-(n-2)(2n-5)+\frac{4b^{2}(2n-3)r_{i}^{2}}{n-1}\right]\nonumber\\
-\frac{64\pi^{2}Q_{c}^{2}}{\omega_{n-1}^{2}} 
- \sqrt{1+\frac{16\pi^{2}Q_{c}^{2}}{b^{2}r_{i}^{2n-2}\omega_{n-1}^{2}}}\left[ \frac{4b^{2}(2n-3)r_{i}^{2n-2}}{n-1}+\frac{64(n-2)\pi^{2}Q_{c}^{2}}{(n-1)\omega_{n-1}^{2}}\right]\nonumber\\
-r_{i}^{2n-4}\left(\frac{nr_{i}^{2}}{l^{2}}-n+2 \right)\left( 1+\frac{16\pi^{2}Q_{c}^{2}}{b^{2}r_{i}^{2n-2}\omega_{n-1}^{2}}\right).   
\end{eqnarray}
Let us now consider the functional relation
\begin{equation}
\Phi=\Phi(r_+,Q)
\end{equation}
from which we find,
\begin{equation}
\left[ \left( \frac{\partial \Phi}{\partial r_+}\right)_{T} \right]_{r_+=r_i} = \left[ \left( \frac{\partial \Phi}{\partial r_+}\right)_{Q} \right]_{r_+=r_i}+\left[ \left( \frac{\partial Q}{\partial r_+}\right)_{T}\right]_{r_+=r_i}\left( \frac{\partial \Phi}{\partial Q}\right)_{r_+=r_i}\label{e1}.
\end{equation}
By using (\ref{Phi}) and (\ref{e2}) we obtain from (\ref{e1}) the result,
\begin{equation}
\left[ \left( \frac{\partial \Phi}{\partial r_+}\right)_{T} \right]_{r_+=r_i} =- \left( \frac{4\pi Q_c}{r_i^{n-1}\omega_{n-1}}\right) \left(1-\frac{8\pi^{2}Q^{2}_{c}}{b^{2}r^{2n-2}_{i}\omega_{n-1}^{2}} \right)\label{phi1}.
\end{equation}  
As a next step, for a fixed value of the temperature ($ T $), we Taylor expand $ \Phi(r_+,T) $ close to the critical point $ r_+=r_i $, which yields,
\begin{eqnarray}
\Phi(r_+)= \Phi(r_i)+\left[ \left( \frac{\partial \Phi}{\partial r_+}\right)_{T=T_i}\right]_{r_+=r_i} (r_+-r_i)
+ higher~~ order~~ terms\label{e3}.
\end{eqnarray}
Ignoring all the higher order terms in (\ref{e3}) and using (\ref{e4}) and (\ref{phi1}) we obtain 
\begin{equation}
\Phi(r_+)-\Phi(r_i)=-\left( \frac{2}{M_i}\right) ^{1/2}\left( \frac{4\pi Q_c}{r_i^{n-1}\omega_{n-1}}\right) \left(1-\frac{8\pi^{2}Q^{2}_{c}}{b^{2}r^{2n-2}_{i}\omega_{n-1}^{2}} \right) |Q-Q_i|^{1/2}\label{e6}.
\end{equation} 
Finally, taking the limit $ b\rightarrow\infty $, we abstract the corresponding expression for the RN AdS black hole as,
\begin{equation}
\Phi(r_+)-\Phi(r_i)=-\left( \frac{2}{[M_i]_{RN AdS}}\right) ^{1/2}\left( \frac{4\pi Q_c}{r_i^{n-1}\omega_{n-1}}\right)|Q-Q_i|^{1/2}\label{rnads4}
\end{equation}
where,
\begin{eqnarray}
[M_i]_{RN AdS}= \frac{(n-1)\omega_{n-1}^{2}[\Gamma(r_i,Q_c)]_{RN AdS}}{64\pi^{2}r_iQ_c}
\end{eqnarray}
with,
\begin{eqnarray}
[\Gamma(r_i,Q_c)]_{RN AdS}=r_{i}^{2n-4}\left[ \frac{n}{l^{2}}(2n-3)r_{i}^{2}-(n-2)(2n-5)+\frac{4b^{2}(2n-3)r_{i}^{2}}{n-1}\right]-\frac{64\pi^{2}Q_{c}^{2}}{\omega_{n-1}^{2}}\nonumber\\ 
- \left[ \frac{4b^{2}(2n-3)r_{i}^{2n-2}}{n-1}+\frac{64(n-2)\pi^{2}Q_{c}^{2}}{(n-1)\omega_{n-1}^{2}}\right]
-r_{i}^{2n-4}\left(\frac{nr_{i}^{2}}{l^{2}}-n+2 \right).
\end{eqnarray}
Comparing (\ref{e6}) and (\ref{rnads4}) with (\ref{PQ}) we note that $ \delta=2 $.

In order to calculate the critical exponent $ \varphi $ we note from (\ref{CQ1}) that near the critical point $ r_+ =r_i$ the heat capacity behaves as
\begin{equation}
C_Q \sim \frac{1}{\Delta}.
\end{equation}
Using (\ref{e4}) this yields our cherished form,
\begin{equation}
C_Q =\frac{A_ir_i\sqrt{M_i}}{\sqrt{2}D_i|Q-Q_i|^{1/2}} \label{e10}.
\end{equation}
For the RN AdS black hole, the corresponding expression becomes,
\begin{equation}
[C_Q]_{RN AdS}=\frac{[A_i]_{RN AdS}r_i\sqrt{[M_i]_{RN AdS}}}{\sqrt{2}[D_i]_{RN AdS}|Q-Q_i|^{1/2}}\label{rnads5}
\end{equation} 
Comparing (\ref{e10}) and (\ref{rnads5}) with the standard relation
\begin{equation}
C_Q \sim \frac{1}{|Q-Q_i|^{\varphi}}.
\end{equation}
we note that $ \varphi=1/2 $.

In an attempt to calculate the critical exponent $ \psi $,  we use (\ref{S}) and (\ref{tri}) in order to expand the entropy ($ S $) near the critical point $ r_+=r_i $. Then it follows, 
\begin{eqnarray}
S(r_+)&=&S(r_i)+\left[ \left( \frac{\partial S}{\partial r_+}\right)\right] _{r_+=r_i}(r_+-r_i)+ higher~~ order~~ terms\nonumber\\ 
&=& S(r_i)+ \frac{\omega_{n-1}}{4}(n-1)r_{i}^{n-1}\Delta \label{e7}
\end{eqnarray}
where we have ignored the higher order terms in $ \Delta $ as we did earlier.
Exploiting (\ref{e4}) we obtain,
\begin{equation}
S(r_+)-S(r_i)=\frac{\omega_{n-1}}{4}(n-1)r_{i}^{n-2}\sqrt{\frac{2}{M_i}}|Q-Q_i|^{1/2}\label{e8}.
\end{equation}
It is now quite straightforward to obtain the corresponding expression for the RN AdS black hole as,
\begin{equation}
S(r_+)-S(r_i)=\frac{\omega_{n-1}}{4}(n-1)r_{i}^{n-2}\sqrt{\frac{2}{[M_i]_{RN AdS}}}|Q-Q_i|^{1/2}\label{rnads6}.
\end{equation}
Comparing (\ref{e8}) and (\ref{rnads6}) with the standard form
\begin{equation}
S(r_+)-S(r_i) \sim |Q-Q_i|^{\psi}
\end{equation}
we note that $ \psi=1/2 $.

Let us now tabulate the various static critical exponents obtained so far.
\begin{table}[h]
\caption{Various static critical exponents and their values}   
\centering                          
\begin{tabular}{c c c c c c c c c}            
\hline\hline                        
$Critical~Exponents  $ & $\alpha$ & $\beta$  & $\gamma$ & $ \delta $ & $\varphi $ & $ \psi $ \\ [2.0ex]
\hline
Values & 1/2 & 1/2 & 1/2 & 2 & 1/2 & 1/2\\ [2.0ex]         
\hline                              
\end{tabular}
\label{E1}          
\end{table}

In usual thermodynamics the critical exponents are found to satisfy certain (scaling) relations among themselves, known as \textit{thermodynamic scaling laws} \cite{stanley1}, which may be expressed as,
\begin{eqnarray}
\alpha + 2\beta +\gamma = 2,~~~\alpha +\beta(\delta + 1) =2,~~~(2-\alpha)(\delta \psi -1)+1 =(1-\alpha)\delta ~~\nonumber\\
~~ \gamma(\delta +1) =(2-\alpha)(\delta -1),~~~\gamma =\beta (\delta - 1),~~~\varphi + 2\psi -\delta^{-1} =1 \label{scaling laws}.
\end{eqnarray}     
From the values of the critical exponents (given in table 5), it is easily seen that all the above relations (\ref{scaling laws}) are indeed satisfied for BI AdS black holes.\\\ \\\\

\textbf{\textit{Static scaling hypothesis:}}

The static scaling hypothesis plays an important role in order to explore the singular behavior of several thermodynamic functions near the critical point. The scaling hypothesis for the static critical phenomena has been made in a variety of situations and applied to thermodynamic functions as well as static and dynamic correlation functions \cite{sh1}-\cite{sh3}. In spite of its several applications to ordinary thermodynamic systems, it is not a widely explored scheme for black holes \cite{cr2},\cite{cr15}. In fact it has never been explored in higher dimensions so far.

For ordinary thermodynamic systems, near the critical point, the free energy may be split into two parts, one regular at the critical point while the other contains the singular part. The scaling hypothesis is made for the singular part. In the subsequent sections we aim to study the scaling hypothesis for the BI AdS black holes in higher dimensions. The corresponding discussion for the RN AdS case arises as a natural consequence of the BI AdS case.  

The static scaling hypothesis for black holes\cite{cr2} states that \textit{close to the critical point, the singular part of the Helmholtz free energy $ F(T,Q)= M-TS $ is a generalized homogeneous function of its variables} i.e; there exists two numbers(known as scaling parameters) $ p $ and $ q $ such that for all positive $ \Lambda $
\begin{equation}
F(\Lambda^{p}\epsilon,\Lambda^{q}\Pi)=\Lambda F(\epsilon,\Pi)\label{ghf}.
\end{equation}

Let us now calculate these scaling parameters for the BI AdS black hole (and consequently for RN AdS case also) in arbitrary dimensions. In order to do that we first Taylor expand $ F(T,Q) $ close to the critical point $ r_+=r_i $ which yields,
\begin{eqnarray}
F(T,Q)= F(T,Q)|_{r_+=r_i} + \left[ \left( \frac{\partial F}{\partial T}\right)_{Q}\right]_{r_+=r_i} (T-T_i)+\left[ \left( \frac{\partial F}{\partial Q}\right)_{T}\right]_{r_+=r_i} (Q-Q_i)\nonumber\\
+\frac{1}{2} \left[ \left( \frac{\partial^{2} F}{\partial T^{2}}\right)_{Q}\right]_{r_+=r_i} (T-T_i)^{2}+\frac{1}{2} \left[ \left( \frac{\partial^{2} F}{\partial Q^{2}}\right)_{T}\right]_{r_+=r_i} (Q-Q_i)^{2} \nonumber\\
+\left[ \left( \frac{\partial^{2}F}{\partial Q \partial T}\right)\right]_{r_+=r_i} (Q-Q_i)(T-T_i) +higher~~~~order ~~~ terms. \label{F1}
\end{eqnarray}
From our previous discussions it is by now quite evident that both 
\begin{equation}
C_Q= - T \left( \frac{\partial^{2} F}{\partial T^{2}}\right)_{Q} ~~~ and ~~~ K^{-1}_{T}= Q \left( \frac{\partial^{2} F}{\partial Q^{2}}\right)_{T}
\end{equation}
diverge at the critical point $ r_+=r_i $. Therefore from (\ref{F1}) we identify the singular part of the free energy as,
\begin{eqnarray}
F_{singular}&=&\frac{1}{2} \left[ \left( \frac{\partial^{2} F}{\partial T^{2}}\right)_{Q}\right]_{r_+=r_i} (T-T_i)^{2}+\frac{1}{2} \left[ \left( \frac{\partial^{2} F}{\partial Q^{2}}\right)_{T}\right]_{r_+=r_i} (Q-Q_i)^{2}\nonumber\\
&=& -\frac{C_Q}{2T_i}(T-T_i)^{2}+\frac{K^{-1}_{T}}{2Q_i} (Q-Q_i)^{2}.
\end{eqnarray} 
Using (\ref{delta}), (\ref{CQ1}), (\ref{KT}) and (\ref{e4}) we finally obtain 
\begin{equation}
F_{singular}= a_i \epsilon^{3/2}+b_i \Pi^{3/2}\label{F}
\end{equation}
where,
\begin{equation}
 a_i=-\frac{A_i T_i }{2}~~~ and, ~~~ b_i=\frac{Q^{3/2}_{i}r_i M^{1/2}_i B_i}{2^{3/2}D^{1/2}_i}.
\end{equation}
Setting the limit $ b\rightarrow\infty $ one can obtain the corresponding expression for RN AdS black hole as,
\begin{equation}
[F_{singular}]_{RN AdS}= [a_i]_{RN AdS} \epsilon^{3/2}+ [b_i]_{RN AdS} \Pi^{3/2}\label{rnads7}
\end{equation}
where,
\begin{equation}
 [a_i]_{RN AdS}=-\frac{[A_i]_{RN AdS} [T_i]_{RN AdS} }{2}~~~ and, ~~~ [b_i]_{RN AdS}=\frac{Q^{3/2}_{i}r_i [M^{1/2}_i]_{RN AdS} [B_i]_{RN AdS}}{2^{3/2}[D^{1/2}_i]_{RN AdS}}.
\end{equation}

Finally, from (\ref{F}) and (\ref{rnads7}) we note that in order to satisfy (\ref{ghf}) we must have $ p=q=2/3 $. Note that although the scaling parameters $ p $ and $ q $ are in general different for a generalized homogeneous function, this is a special case  where both the numbers $ p $ and $ q $ have identical values. In other words $ F $ behaves as a usual homogeneous function.

At this stage it is important to note that, in standard thermodynamics, the various critical exponents are related to the scaling parameters as, \cite{stanley1}
\begin{eqnarray}
\alpha = 2-\frac{1}{p},~~~\beta =\frac{1-q}{p},~~~\delta =\frac{q}{1-q}~~\nonumber\\
~~ \gamma =\frac{2q-1}{p},~~~\psi =\frac{1-p}{q},~~~\varphi =\frac{2p-1}{q}
\end{eqnarray}
It is reassuring to note that these relations are also satisfied both for the BI AdS and RN AdS black holes. One can further see that elimination of the two scaling parameters ($ p $ and $ q $) from the above set of relations eventually leads to (\ref{scaling laws}). This is consistent with the fact that the scaling hypothesis replaces the exponent \textit{inequalities} with the \textit{equalities} \cite{stanley1}-\cite{stanley2}, as found in usual systems. The fact that all the critical exponents can be expressed in terms of two scaling parameters ($ p $ and $ q $) suggests that if two exponents are specified then all others can be determined. 

Finally, we are in a position to provide a suggestive way to find out the rest of the critical exponents $ \nu $ and $ \eta $ which are associated with the  correlation length and correlation function respectively. Assuming the additional scaling relations \cite{stanley1},
\begin{equation}
\gamma = \nu (2-\eta),~~~ 2-\alpha = \nu d \label{add}
\end{equation}
to be valid, where $ d(=n) $ is the spatial dimensionality of the system, we find  
\begin{equation}
\nu =\frac{3}{2n},~~~ \eta =\frac{6-n}{3}.
\end{equation}
Frankly speaking, till date it is not clear a priory whether these additional relations (\ref{add}) are indeed valid for black holes. It will be more definitive if one attempts to compute the values for $ \nu $ and $ \eta $ directly from the correlation of scalar modes in gravity theory \cite{cr9}. 

\section{Conclusions}
In this paper, based on a canonical framework, we have provided an approach to study the issue of critical phenomena for charged black holes in higher dimensional AdS space. A number of interesting features have been observed in this regard. First of all we note that the boundedness on the parameter space ($ b $ and $ Q $)that exists in ($3+1$) dimension\cite{myung} eventually disappears in higher dimensions. Furthermore, based on thermodynamic framework we have systematically addressed  various aspects of the critical phenomena in higher dimensional Born-Infeld AdS (BI AdS) black holes. Considering the appropriate limit ($ b\rightarrow\infty $) the critical behavior of higher dimensional Reissner Nordstrom AdS (RN AdS) black holes has also been studied simultaneously. Based on a thermodynamic approach we have calculated all the static critical exponents ($\alpha=1/2,\beta=1/2,\gamma=1/2,\delta=2,\varphi=1/2,\psi=1/2$) both for the BI AdS and RN AdS black holes. They are also found to satisfy the so called  thermodynamic scaling relations (\ref{scaling laws}) near the critical point which is characterized by the discontinuity in the heat capacity at constant charge ($ C_Q $).  We have also explored the static scaling hypothesis which has been found to be compatible with the scaling relations near the critical point. The scaling parameters have been found to possess identical values ($ p=q=2/3 $). It is quite remarkable to note that the scaling parameters do not depend on the spatial dimensionality of the AdS space.  Moreover we have checked the additional scaling relations in order to gain some insights regarding the critical exponents ($ \nu=\frac{3}{2n},\eta=\frac{6-n}{3} $) associated with the spatial correlation. It is quite interesting to note that the static critical exponents (table 5) thus obtained are independent of the spatial dimensionality of the AdS space time. This indeed suggests the \textit{mean field} behavior in black holes as  thermodynamic system and bolsters our confidence in applying Ehrenfest's scheme to study phase transitions \cite{bss}-\cite{dibakar2}. On top of it, from our analysis the thermodynamic scaling relations for black holes are found to be valid in any dimensions. This is also compatible with a mean field analysis that illustrates the dimension independence of the static critical exponents in usual thermodynamic systems. As a matter of fact, from our analysis we find that both BI AdS and RN AdS black holes posses identical critical exponents and thereby similar scaling behavior near the critical point. This indeed suggests that both these types of black holes belong to the same universality class. It is quite reassuring to note that the present study of critical phenomena in black holes vindicates the mean field approximation that generally holds for long range interaction.

Although we have resolved a number of vexing issues regarding the critical behavior of charged AdS black holes in arbitrary dimensions, still there remain a few more points that admit a further investigation into the subject. For example in order to calculate the critical exponents $ \nu $ and $ \eta $ we have assumed the validity of the additional scaling relations (\ref{add}), which may not be true in practice. In fact this is a highly nontrivial issue in higher dimensions. Therefore as an alternative approach one should compute them from a knowledge of correlation scalar modes in the BI AdS back ground which would further clarify our calculations. It would also be interesting to exploit the AdS/CFT duality in order to gain a new insight to the subject. It would be an interesting issue to find out whether there exists any duality between the critical behavior of the black holes in the bulk to that with the boundary CFTs. Also, the investigation of the underlying renormalization group structure could be an alternative way to explain the critical behavior of black holes.

{\bf{ Acknowledgement:}}\\
 D.R would like to thank the Council of Scientific and Industrial Research (C. S. I. R), Government of India, for financial help.

  

\end{document}